\documentclass[12pt]{article}
\usepackage[affil-it]{authblk}
\usepackage{natbib}
\usepackage{amsmath}
\usepackage{amsfonts}
\usepackage{bm}
\usepackage[color=lightgray]{todonotes}
\usepackage[bf,font={small,sl}]{caption}
\usepackage{booktabs}
\usepackage[margin=1in]{geometry}

\title{\textbf{Linking resource selection and step selection models for habitat preferences in animals}}
\author{Th\'eo Michelot$^1$\footnote{Corresponding author: tmichelot1@sheffield.ac.uk}, Paul G. Blackwell$^1$, Jason Matthiopoulos$^2$}
\affil{$^1$University of Sheffield, $^2$University of Glasgow}
\date{}

\begin{document}
\maketitle

\begin{abstract}
\emph{The two dominant approaches for the analysis of species-habitat associations in animals have been shown to reach divergent conclusions. Models fitted from the viewpoint of an individual (step selection functions), once scaled up, do not agree with models fitted from a population viewpoint (resource selection functions). We explain this fundamental incompatibility, and propose a solution by introducing to the animal movement field a novel use for the well-known family of Markov chain Monte Carlo (MCMC) algorithms. By design, the step selection rules of MCMC lead to a steady-state distribution that coincides with a given underlying function: the target distribution. We therefore propose an analogy between the movements of an animal and the movements of a MCMC sampler, to guarantee convergence of the step selection rules to the parameters underlying the population's utilisation distribution. We introduce a rejection-free MCMC algorithm, the local Gibbs sampler, that better resembles real animal movement, and discuss the wide range of biological assumptions that it can accommodate. We illustrate our method with simulations on a known utilisation distribution, and show theoretically and empirically that locations simulated from the local Gibbs sampler give rise to the correct resource selection function. Using simulated data, we demonstrate how this framework can be used to estimate resource selection and movement parameters.}
\end{abstract}

\vspace{1em}
\vspace{1em}

\newpage
\section{Introduction}
Understanding how animals use a landscape in response to its habitat composition is a crucial question in pure and applied ecology. Such insights are achievable only by confronting species-habitat association models with usage data, collected either via transect surveys or via biologging methods. Statistical inference, to link these data to environmental variables, can be approached from a population perspective, using resource selection functions \citep[RSF;][]{manly2007resource}. Alternatively, if individually referenced data (i.e.\ telemetry) are available, the question can be addressed from the viewpoint of the single animal, via step selection functions \citep[SSF;][]{thurfjell2014applications}. The population/individual dichotomy between these two approaches is not always clear-cut, because RSFs can be applied to the utilisation distribution of single animals, and SSFs can combine joint insights from multiple individuals. Nevertheless, the two methods roughly fall at opposite ends of the Eulerian-Lagrangian spectrum outlined by \cite{turchin1998quantitative}. Therefore, researchers in this area have tended to think of the habitat preference parameters obtained via SSFs as the microscopic rules of movement, while the corresponding parameters of an RSF are implicitly thought of as the macroscopic patterns obtained in the long term. Hence, SSF models are increasingly concerned with the geometry of movement trajectories \citep[e.g.\ step lengths and turning angles in different behavioural states in][]{squires2013combining}, while RSF predictions often make a pseudo-equilibrium assumption \citep{guisan2005predicting}, which is a biological term reminiscent of the mathematical idea of steady-state distributions. But herein lies a fundamental problem for this entire field of statistical analysis. A correctly formulated framework of movement must work across scales, such that, when the microscopic rules of individual movement are scaled up in space and time, they give rise to the expected macroscopic distribution of a population. However, there is now both analytical \citep{barnett2008analytic,moorcroft2008mechanistic} and numerical \citep{signer2017estimating} evidence that the distribution constructed from the coefficients of a SSF does not match the spatial predictions of the RSF fitted to the same data. Here, we explain how this discrepancy arises and propose a solution.

 A RSF $w(\bm{c})$ is proportional to the probability of a unit of habitat $\bm{c}$ being used \citep{boyce1999relating}. Depending on the type of usage data available, RSFs are derived in two steps. First, a model is fitted to the response and explanatory data. For example, a point process model \citep{aarts2012comparative} or a use-availability logistic regression \citep{boyce1999relating, aarts2008estimating} can be used for telemetry data, and a log-linear regression can be used on count data from regular grids or line transects. Second, irrespective of the type of response data and model fitting method, the linear predictor of the resulting statistical model is transformed via a non-negative function \citep[][Chapter 2]{manly2007resource}, of which the most common is the exponential,
\begin{equation} \label{eqn:rsf}
  w(\bm{c}) = \exp (\beta_1 c_1 + \beta_2 c_2 + \dots + \beta_m c_m),
\end{equation}
where $\bm{c}$ is a vector of $m$ covariate values, and $\beta_1, \beta_2, \dots, \beta_m$ are the associated regression coefficients. The RSF can be used to model the utilisation distribution $\pi(\bm{x})$, i.e.\ the distribution of the animal's space use,
\begin{equation} \label{eqn:ud}
	\pi(\bm{x}) = \dfrac{\exp (\beta_1 c_1(\bm{x}) + \beta_2 c_2(\bm{x}) + \dots + \beta_m c_m(\bm{x}))}
    	{\int_\Omega \exp (\beta_1 c_1(\bm{z}) + \beta_2 c_2(\bm{z}) + \dots + \beta_m c_m(\bm{z})) d\bm{z}},
\end{equation}
where the functions $c_1, c_2, \dots, c_m$ associate a spatial location $\bm{x}$ to the corresponding covariate values, and $\Omega$ is the study region. The utilisation distribution is normalized by the denominator in Equation \ref{eqn:ud} to ensure that it defines a valid probability distribution for $\bm{x}$, hence the lack of an intercept in the linear predictor. Although they can encompass a wider range of environmental conditions, the covariates are often called resources in this context. In the following, we use ``covariates'' and ``resources'' interchangeably.

RSF approaches are commonly used to estimate the apparent effect of a spatial covariate on a species. The resource selection coefficients $\beta_k$ characterize this effect for each of the $m$ covariates ($\beta_k>0$: preference; $\beta_k<0$: avoidance; $\beta_k=0$: indifference; see \cite{avgar2017relative} for a discussion of the interpretation of the $\beta_k$ in terms of selection strength). However, recent work has shown that these interpretations are highly sensitive to the context in which the organisms are being studied, in particular, the availability of all habitat types to the animals \citep{beyer2010interpretation, matthiopoulos2011generalized, paton2016defining}. Thus, in this framework, the definition of habitat availability, determined by assumptions of spatial accessibility \citep{matthiopoulos2003use}, is important in deducing preference from observed usage. For example, when using RSFs to analyse a time series of positions from a ranging animal, it may not be plausible to assume that all locations in the home range are accessible by the animal at every step \citep{northrup2013practical}. RSF approaches are often forced to treat such non-independence as a statistical nuisance \citep{aarts2008estimating, fieberg2010correlation, johnson2013estimating}, but step selection approaches treat it as an asset.

In step selection analyses, the likelihood $p(\bm{y} \vert \bm{x})$ of a potential displacement by the animal to a location $\bm{y}$ over a given time interval (typically, the sampling interval) is modelled in terms of the habitat composition in the neighbourhood of the animal's current position $\bm{x}$:

\begin{equation} \label{eqn:ssf}
  p(\bm{y} \vert \bm{x}) = \dfrac{\phi(\bm{y} \vert \bm{x}) w(c(\bm{y}))}
  {\int_\Omega \phi(\bm{z} \vert \bm{x}) w(c(\bm{z})) d\bm{z}},
\end{equation}
where $\phi(\cdot \vert \bm{x})$ is defined over a spatial domain $\Omega$, and, for any location $\bm{x}$, $$c(\bm{x}) = (c_1(\bm{x}),c_2(\bm{x}),\dots,c_m(\bm{x})).$$ The function $\phi(\cdot \vert \bm{x})$ is called the resource-independent movement kernel around $\bm{x}$ \citep{rhodes2005spatially, forester2009accounting}, and it describes the density of endpoints for a step starting in $\bm{x}$, in the absence of resource selection. To link the movement to environmental covariates, $w$ is modelled using the same log-linear link as the RSF, given in Equation \ref{eqn:rsf}. In this context, the term ``step selection function'' (SSF) is most often used for $w$ \citep[e.g.\ by][]{fortin2005wolves, thurfjell2014applications}; however, note that it is sometimes used for the whole numerator in the right-hand side of Equation \ref{eqn:ssf} \citep[see][]{forester2009accounting}. In the following, we call $w$ the SSF.

The choice of the function $\phi$ characterizes accessibility, and hence determines availability, in a step selection model; it corresponds to the distribution of feasible steps over one time interval, with origin $\bm{x}$, when the resources do not affect the movement. It can, for example, be a uniform distribution on a disc around the current location $\bm{x}$ \citep[e.g.][]{arthur1996assessing}, or obtained from the empirical distributions of movement metrics \citep[e.g.\ step lengths and turning angles in][]{fortin2005wolves}. 

SSFs are most often fitted using conditional logistic regression on matched use-availability data, where each observed step $\bm{x}_t \rightarrow \bm{x}_{t+1}$ is matched to a set of random steps generated from $\phi(\cdot \vert \bm{x}_t)$ \citep{thurfjell2014applications}. \cite{duchesne2015equivalence} showed that a step selection model defines a movement model equivalent to a biased correlated random walk (BCRW). BCRWs are routinely used in ecology as a flexible basis for models of individual movement \citep{turchin1998quantitative, codling2008random}. \cite{avgar2016integrated} extended the step selection approach to allow simultaneous inference on habitat selection and on the movement process, making it a very attractive framework to estimate habitat preference from movement data \citep{prokopenko2017characterizing, scrafford2018roads}. Step selection models have been used to analyse the impact of landscape features on animal space use \citep[e.g.][]{coulon2008inferring, roever2010grizzly}, as well as animal interactions \citep{potts2014unifying}.

Although the RSF and SSF are typically described with the same notation, and used for the same purpose of estimating habitat preference, it can be shown that their steady-state predictions do not generally coincide. For a known utilisation distribution, \cite{signer2017estimating} showed empirically that the normalized SSF (``naive'' estimate) differed from the utilisation distribution. In particular, the difference was greater when $\phi$ was narrow compared to the scale of habitat features. Similarly, \cite{barnett2008analytic} showed that, for the step selection model defined in Equation \ref{eqn:ssf}, the steady-state distribution of the animal's location (i.e.\ its utilisation distribution) is given by

\begin{equation}\label{eqn:steadystate}
  \pi(\bm{x}) = \dfrac{w(c(\bm{x})) \int w(c(\bm{y})) \phi(\bm{y} \vert \bm{x}) d\bm{y}}
  					{\int w(c(\bm{y})) \int w(c(\bm{z})) \phi(\bm{z} \vert \bm{y}) d\bm{z} d\bm{y}}.
\end{equation}

That is, the steady-state distribution of the model is generally not proportional to the SSF $w$, and that discrepancy crucially depends on the choice of the resource-independent movement kernel $\phi$. An example of this is their earlier result \citep{moorcroft2008mechanistic} that under one specific set of assumptions, the steady-state distribution is approximately proportional to the \emph{square} of the SSF.

Although it may seem disconcerting that the two approaches lead to different estimates of $w$, the cause of this apparent paradox is partly due to the notational misuse of the same symbol for what are, in effect, different objects. The SSF captures local aspects of the animal's movement, because it only considers a neighbourhood of the current location of the animal (determined by $\phi$) and only becomes a better approximation of the RSF when the scale of $\phi$ increases \citep{barnett2008analytic}. The parameters of the two objects coincide in the limiting case of unconstrained mobility, i.e.\ when the availability assumed by both methods is global. However, in every other case, the two methods are different. \cite{schlagel2016robustness} also noted that, unlike RSF models, standard SSFs are scale-dependent, in that their habitat selection estimates depend on the time scale of the observations \citep[although see][for a SSF approach with a user-defined scale of selection]{hooten2014temporal}.

Several approaches have been suggested to approximate the steady-state distribution of SSF movement models. In particular, \cite{avgar2016integrated} and \cite{signer2017estimating} showed that simulations from a fitted SSF could be used to obtain estimates of the underlying utilisation distribution. Similarly, \cite{potts2014predicting} described a numerical method to compute the utilisation distribution given in Equation \ref{eqn:steadystate}, as it generally has no closed form expression. Those approaches are useful to predict space use from SSFs, but they do not allow the steady-state distribution of locations to be modelled in a simple parametric form, as in Equation \ref{eqn:ud}. One important consequence is that, because the utilisation distribution of SSF models is not modelled by a RSF, joint inference from telemetry data and survey data into habitat selection and space use has not been possible with existing approaches.

Rather than seeking an equivalence of the parameters estimated by RSF and SSF methods, a better question to ask is: under what assumptions do the parameters estimated by a SSF lead to movement that scales to the distribution yielded by the parameters of a RSF model? In Section \ref{sec:method}, we reconcile resource selection and step selection conceptually, with a new step selection model for which the long-term distribution of locations is guaranteed to be proportional to the RSF. Our method uses an analogy between the movement of an animal in geographical space and the movement of a Markov chain Monte Carlo (MCMC) sampler in its parameter space. In Section \ref{sec:gibbs}, we make these concepts applicable in practice, by developing a family of MCMC algorithms with considerable potential for encompassing realistic movement assumptions. In Section  \ref{sec:sim}, we illustrate our method using simulations on a known utilisation distribution. We verify that the distribution of simulated locations corresponds to the correct RSF, and we present a proof-of-concept analysis to demonstrate the potential of the method for estimating resource selection coefficients and parameters of the movement process from telemetry data.

\section{A model of step selection using a movement-MCMC analogy}
\label{sec:method}

MCMC methods are a general framework to sample from a probability distribution, termed the target distribution \citep{gilks1995markov}. This approach is mostly used for Bayesian inference, to sample from the (posterior) distribution of a set of unknown parameters \citep[][Chapter 11]{gelman2014bayesian}. It includes a very wide class of algorithms, among them the widely-used Metropolis-Hastings and Gibbs samplers. A MCMC algorithm describes the steps to generate a sequence of points $\bm{x}_1,\bm{x}_2,\bm{x}_3\dots$, whose long-term distribution is the target distribution. Each MCMC algorithm is defined by its transition kernel $p(\bm{x}_{t+1} \vert \bm{x}_t)$, which determines (for any $t=1,2,\dots$) how the point $\bm{x}_{t+1}$ should be sampled, given $\bm{x}_t$. For example, in a Metropolis-Hastings algorithm, the transition kernel is a combination of the proposal distribution and the acceptance probability:
\begin{equation*}
  p(\bm{x}_{t+1} \vert \bm{x}_t) = p(\bm{x}_{t+1} \text{ is proposed } \vert\ \bm{x}_t)~ 
  p(\bm{x}_{t+1} \text{ is accepted } \vert\ \bm{x}_t).
\end{equation*}

In general, given some easily-satisfied technical conditions, 
a sufficient condition for $p(\bm{x}_{t+1} \vert \bm{x}_t)$ to define a valid MCMC algorithm for the target distribution $\pi$ (i.e.\ to ensure that the distribution of samples will converge to $\pi$) is the detailed balance condition:
\begin{equation} \label{eqn:DBC}
  \forall \bm{x}, \bm{y},\qquad \pi(\bm{y}) p(\bm{x} \vert \bm{y}) = \pi(\bm{x}) p(\bm{y} \vert \bm{x}).
\end{equation}
That is, if the process is in equilibrium with distribution $\pi$, then the rates of moves in each direction between any $\bm{x}$ and $\bm{y}$  balance out.

We propose an analogy between an animal's observed movement in $n$-dimensional geographical space, and the movement of a MCMC sampler in a $n$-dimensional parameter space, for which the target distribution is the utilisation distribution. That is, we consider that a tracked animal ``samples'' spatial locations in the short term from some movement model, and in the long run from its utilisation distribution, in the same way that a MCMC algorithm samples points in the short term from some transition kernel and in the long term from its target distribution. A MCMC algorithm then defines a movement model, for which the steady-state distribution is known. The dynamics of the movement process $(\bm{x}_t)$ are described by the transition kernel of the algorithm such that, at each time point $t=1,2,\dots$, the next location $\bm{x}_{t+1}$ is sampled from $p(\bm{x}_{t+1} \vert \bm{x}_t)$. By the properties of MCMC samplers, the steady-state distribution for $\bm{x}_t$ is $\pi$. The utilisation distribution can be modelled with the RSF, as defined in Equation \ref{eqn:ud}, to link the target distribution of the movement model to the distribution of resources.

A MCMC algorithm, if viewed as a movement model, can then be used to analyse animal tracking data, in the following steps. Although we focus on step 1 in this paper, we illustrate steps 2 and 3 with a simulated example in Section \ref{sec:estim}. 
\begin{enumerate}
\item Choose a MCMC algorithm, to be used as a model of animal movement and habitat selection. We suggest one such algorithm in Section \ref{sec:gibbs}.
\item Write the likelihood of the model. Under a MCMC movement model, the likelihood of an observed step from $\bm{x}_t$ to $\bm{x}_{t+1}$ is a function of the resource selection coefficients and of the other parameters of the sampler, given by the transition kernel $p(\bm{x}_{t+1} \vert \bm{x}_t)$.
\item Use maximum likelihood estimation, or other likelihood-based methods, to estimate the resource selection and movement parameters.
\end{enumerate}

In this framework, the choice of the MCMC algorithm determines the movement model. For example, with a Metropolis-Hastings model, different proposal distributions might capture different features of the animal's movement. The parameters of the algorithm, which are usually regarded as tuning parameters, are here parameters of the movement process. For example, the variance of the proposal distribution can be thought of as a measure of the animal's speed. It is important to make a distinction between these parameters of movement, and the parameters of the target distribution (i.e.\ the resource selection parameters). Two different samplers might have the same target distribution, but the rate at which it is approached by the MCMC samples will depend on the choice of algorithm. Indeed, part of the success of MCMC in its Bayesian context is the flexibility in choosing the transition kernel for a given target distribution. The suitability of a MCMC sampler is usually assessed by the speed of convergence of the simulated samples to the target distribution. However, for our application, we want an algorithm corresponding to a realistic model of movement, in addition to having the correct target distribution. It could happen that a MCMC algorithm that describes animal movement very realistically has a slow rate of convergence to the target distribution. This would merely mean that the animal, when observed at the time step of the observations, does not sample efficiently from its utilisation distribution. In such a case, inference about the utilisation distribution would be limited regardless of the modelling framework that is used.

In rejection-based MCMC algorithms such as Metropolis-Hastings, a relocation is proposed at each time step, and is accepted with some probability. If the proposed step is not accepted, the process remains in the same location. Although it can happen that a tagged animal is immobile over several time steps (in particular if temporal resolution is high), many telemetry data sets do not include such ``rejections''. Classic MCMC algorithms might thus seem to be an unnatural choice to analyse those data, because the animal will almost always change position in the process of sampling a new candidate location. To circumvent this problem, we design a new rejection-free MCMC algorithm in Section \ref{sec:gibbs}.

\section{The local Gibbs sampler}
\label{sec:gibbs}
Standard Metropolis-Hastings samplers require a rejection step to ensure convergence to the target distribution. Viewing this as a movement model would imply the unlikely scenario of a return by the animal to its previous position, after having tested and rejected a relocation. 
Instead, it is more natural to think about tracking data as the outcome of a rejection-free sampler. Several such algorithms are possible; see the Discussion. Here, we describe one such algorithm, which we call the local Gibbs sampler.

In the classic Gibbs sampler, each `step' involves updating just one of the $n$ parameters, $x_j$ say, while keeping $x_1,\ldots, x_{j-1}, x_{j+1}, \ldots,x_{n}$ fixed; the values of $j$ can be chosen systematically or randomly. Thus, each step is a move within a one-dimensional subspace of the parameter space, rather than over the whole space. It is used when the target distribution over each such one-dimensional space (the so-called `full conditional distribution') is mathematically tractable, so that when it is used as the transition kernel for that step, the acceptance probability is guaranteed to be 1.

The local Gibbs sampler uses the same idea of sampling from a restricted part of the target distribution: at each iteration $t$, the updated parameter $\bm{x}_{t+1}$ is sampled directly from the target distribution, truncated to some neighbourhood of $\bm{x}_t$. The way in which this neighbourhood is selected is crucial to ensuring that the algorithm samples from the required target distribution in the long run. 

In explaining the details of the algorithm, we focus on the case of $n=2$ dimensions, by far the most important case for ecological applications, though the algorithm works for any $n$ with straightforward changes. For any point $\bm{x}$, and $r > 0$, we define $\mathcal{D}_r(\bm{x})$ to be the disc of centre $\bm{x}$ and radius $r$.

\begin{figure}[htbp]
  \centering
  \includegraphics[width=0.5\textwidth]{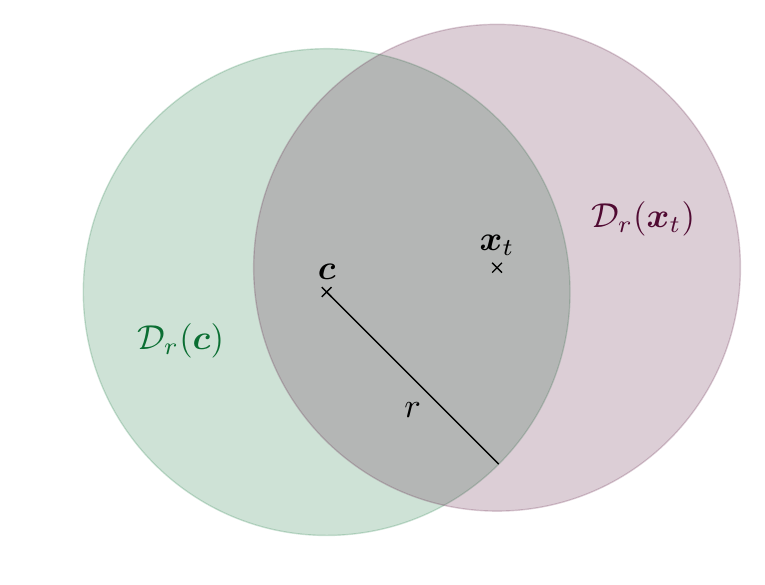}
  \caption{Notation for the local Gibbs sampler in two dimensions. The point $\bm{c}$ is sampled uniformly from $\mathcal{D}_r(\bm{x}_t)$, and the next location $\bm{x}_{t+1}$ is sampled from the RSF truncated to $\mathcal{D}_r (\bm{c})$.}
  \label{fig:gibbsnotation}
\end{figure}

The local Gibbs sampler for $\pi$ is given by the following steps, and the notation is illustrated in Figure \ref{fig:gibbsnotation}. The track starts from a location $\bm{x}_1$, and moves to locations $\bm{x}_{t+1}$ over iterations $t=1,2,\dots$.
\begin{enumerate}
\item On iteration $t$, sample a point $\bm{c}$ uniformly from the disc $\mathcal{D}_r(\bm{x}_t)$.
\item Define $\tilde\pi$ the truncated distribution,
  \begin{equation*}
    \tilde\pi (\bm{y}) =
    \begin{cases}
      \pi(\bm{y})/C_r(\bm{c}) & \text{if } \bm{y} \in \mathcal{D}_r(\bm{c}), \\
      0 & \text{elsewhere,}
    \end{cases}
  \end{equation*}
  where $C_r(\bm{c}) = \int_{\bm{z} \in \mathcal{D}_r(\bm{c})} \pi(\bm{z}) d\bm{z}$ is a normalizing constant.
\item Sample the next location $\bm{x}_{t+1}$ from $\mathcal{D}_r(\bm{c})$ according to the constrained pdf $\tilde\pi$.
\end{enumerate}

The local Gibbs sampler has one parameter: the radius $r>0$ of the relocation disc. Here, for simplicity, we only consider the case where $r$ is fixed, but the algorithm would still work if $r$ were generated independently at each iteration from a probability distribution. 

Using the analogy introduced in Section \ref{sec:method} between animal movement and MCMC sampling, the local Gibbs algorithm can be used as the basis for a model of animal movement and habitat selection, that we will call the local Gibbs model. It relies on the assumption that an animal ``samples'' locations from its utilisation distribution based on the step selection rules described above. 

Note that, at each time step, the overall relocation region of the local Gibbs model is symmetric around the animal's current location. The choice of the relocation disc $\mathcal{D}_r(\bm{c})$, based on the selection of a point $\bm{c}$ in step 1 of the algorithm, might seem biologically unrealistic, because a moving animal would not relocate to a disc that is shifted at random from its current location. Nevertheless, because $\bm{c}$ is chosen uniformly from $\mathcal{D}_r(\bm{x}_t)$, one should think of the relocation region once $\bm{c}$ has been integrated over, i.e.\ a disc of radius $2r$ around $\bm{x}_t$.

In the local Gibbs model, the parameter $r$ determines the size of the area that is available to the animal over one time step. As in most step selection analyses, the region of availability is a simplistic but useful model for a combination of the animal's mobility and perception.

Taking $\pi$ to be the normalized RSF (Equation \ref{eqn:ud}), the local Gibbs algorithm defines a step selection (movement) model in which the distribution of the animal's space use is guaranteed to be proportional to the RSF. Indeed, it satisfies the detailed balance condition (Equation \ref{eqn:DBC}), which can be shown as follows. Given $r$, we have
\begin{align*}
  \pi(\bm{x})p(\bm{y}|\bm{x}) 
  & = \pi(\bm{x}) \int_{\bm{c}\in \mathbb{R}^2} p(\bm{y}|\bm{c})p(\bm{c}|\bm{x}) d\bm{c}
\end{align*}

Given $\bm{c}$, $\bm{y}$ is sampled from $\mathcal{D}_r(\bm{c})$ with a density proportional to $\pi(\bm{y})$ and, given $\bm{x}$, $\bm{c}$ is sampled uniformly from $\mathcal{D}_r(\bm{x})$, so

\begin{equation*}
  p(\bm{y} \vert \bm{c}) = \dfrac{\pi(\bm{y})}{C_r(\bm{c})} I_{\{\bm{y} \in \mathcal{D}_r(\bm{c})\}},
  \text{ and }
  p(\bm{c} \vert \bm{x}) = \dfrac{1}{\pi r^2} I_{\{\bm{c} \in \mathcal{D}_r(\bm{x})\}},
\end{equation*}
where $I_{A}$ is the indicator function for the event $A$. We can then write

\begin{align*}
  \pi(\bm{x})p(\bm{y}|\bm{x}) & = \pi(\bm{x}) 
                                \int_{\bm{c} \in \mathcal{D}_r(\bm{x}) \cap \mathcal{D}_r(\bm{y})} 
                                \dfrac{\pi(\bm{y})}{\pi r^2 C_r(\bm{c})} d\bm{c}\\
                              & = \dfrac{\pi(\bm{x}) \pi(\bm{y})}{\pi r^2}
                                \int_{\bm{c} \in \mathcal{D}_r(\bm{x}) \cap \mathcal{D}_r(\bm{y})} \dfrac{1}{C_r(\bm{c})} d\bm{c} \\
                              & = \dfrac{\pi(\bm{y}) \pi(\bm{x})}{\pi r^2}
                                \int_{\bm{c} \in \mathcal{D}_r(\bm{y}) \cap \mathcal{D}_r(\bm{x})} \dfrac{1}{C_r(\bm{c})} d\bm{c} \\
                              & = \pi(\bm{y})p(\bm{x}|\bm{y}),
\end{align*}
as required.

The local Gibbs model is superficially similar to the availability radius model of \cite{rhodes2005spatially}, first introduced by \cite{arthur1996assessing}. In that model, at each time step, the next location $\bm{x}_{t+1}$ is sampled from the RSF truncated and scaled on a disc centred on $\bm{x}_t$. That is, in step 1 of the algorithm described above, they take $\bm{c}=\bm{x}_t$. This means that there is no mechanism in their approach to guarantee that the overall distribution of the sampled locations is the RSF. Specifically, the two sides of the detailed balance equation involve different normalization constants, and so their movement models do not have the normalized RSF as their equilibrium distributions.
For this reason, the coefficients they estimate will differ from the resource selection coefficients estimated from a RSF approach.

\begin{figure}[hbtp]
  \centering
  \includegraphics[width=\textwidth]{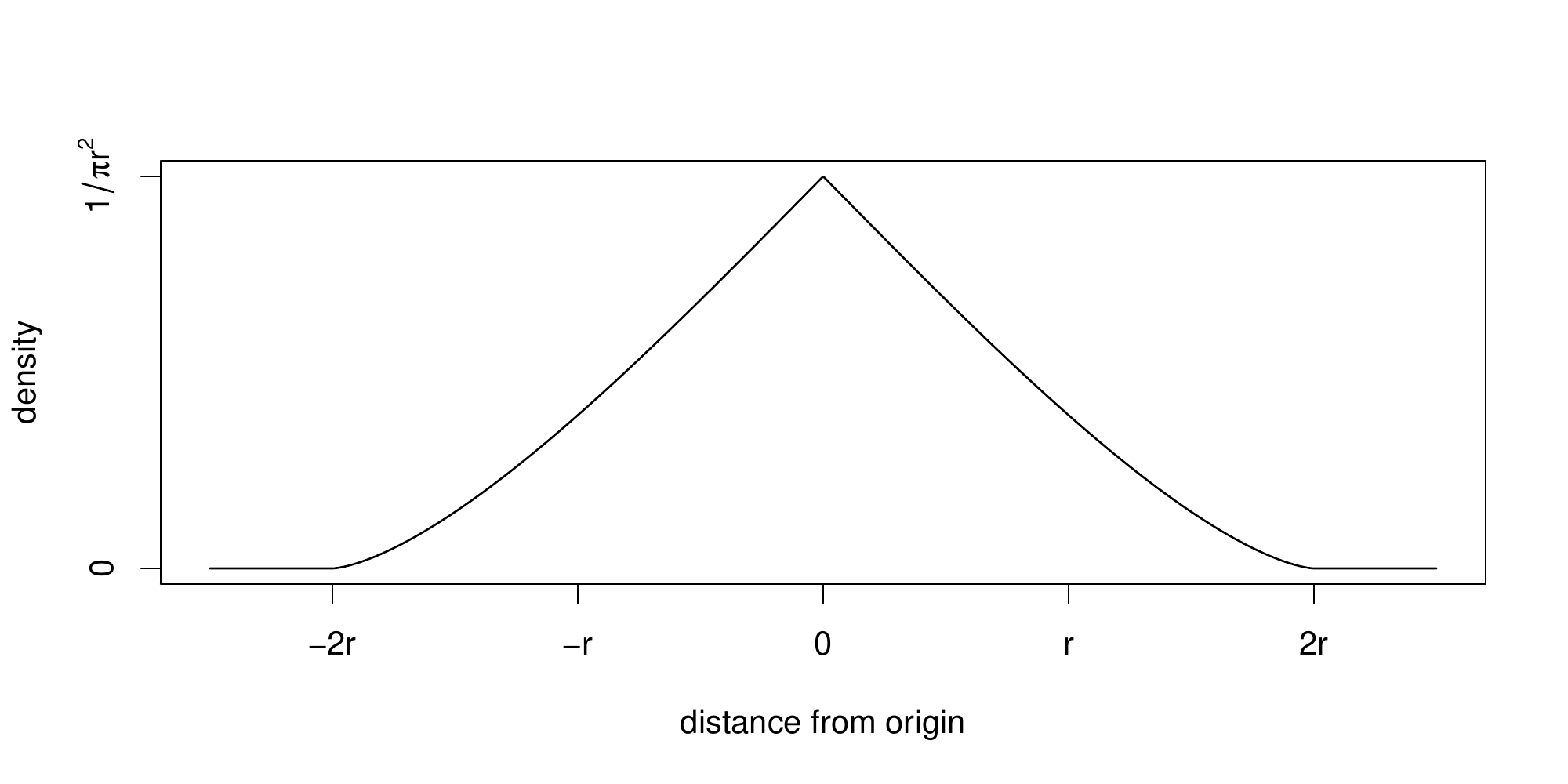}
  \caption{Resource-independent transition kernel for the local Gibbs sampler with a fixed radius parameter $r$. The x axis shows the distance from the origin point $\bm{x}_t$, and the y axis shows the density of the endpoint $\bm{x}_{t+1}$.}
  \label{fig:density}
\end{figure}

We can derive the resource-independent movement kernel $\phi_\text{LG}(\bm{y} \vert \bm{x})$ of the local Gibbs model, to describe the distribution of steps on a flat target distribution. In the case where $r$ is fixed,
\begin{equation} \label{eqn:kernel}
  \phi_\text{LG}(\bm{y} \vert \bm{x}) = 
  \begin{cases}
    \dfrac{1}{(\pi r^2)^2} \mathcal{A}(\mathcal{D}_r(\bm{x}) \cap \mathcal{D}_r(\bm{y})) & \text{if } \lVert \bm{y} - \bm{x} \rVert \leq 2r,\\[2mm]
    0 & \text{otherwise},
  \end{cases} 
\end{equation}
where $\lVert \bm{y} - \bm{x} \rVert$ is the distance between $\bm{x}$ and $\bm{y}$, and $\mathcal{A}(\mathcal{D}_r(\bm{x}) \cap \mathcal{D}_r(\bm{y}))$ is the area of the intersection of the discs of centres $\bm{x}$ and $\bm{y}$, and of radius $r$. The point $\bm{c}$ is such that $\lVert \bm{c} - \bm{x} \rVert < r$ and $\lVert \bm{c} - \bm{y} \rVert < r$, and so -- in the absence of environmental effects -- the relative probability of a step from $\bm{x}$ to $\bm{y}$ is proportional to $\mathcal{A}(\mathcal{D}_r(\bm{x}) \cap \mathcal{D}_r(\bm{y}))$. By construction, it is impossible to have a step between two points if the distance between them is larger than $2r$, hence $\phi_\text{LG}(\bm{y} \vert \bm{x}) = 0$ when $\lVert \bm{y} - \bm{x} \rVert > 2r$. The detail of the derivation is given in Appendix S1. A graph of the density function $\phi_\text{LG}$ is shown in Figure \ref{fig:density}.

The transition kernel given in Equation \ref{eqn:kernel} and plotted in Figure \ref{fig:density} describes the distribution of steps in the absence of habitat selection, in the case where the radius parameter $r$ is fixed. A more flexible movement model can be obtained by taking $r$ to be time-varying, and drawn at each time step from a probability distribution (e.g.\ exponential or gamma distribution, to ensure $r>0$).

It is important to note that the transition kernel of the local Gibbs algorithm cannot be written in the form given in Equation \ref{eqn:ssf}, i.e.\ $p(\bm{y} \vert \bm{x})$ is in general \emph{not} proportional to $\phi_\text{LG}(\bm{y}\vert\bm{x}) w(c(\bm{y}))$. For this reason, the local Gibbs model is not merely a special case of the step selection model described by \cite{forester2009accounting}. 

\section{Simulations}
\label{sec:sim}
\begin{figure}[htbp]
  \centering
  \includegraphics[width=0.6\textwidth]{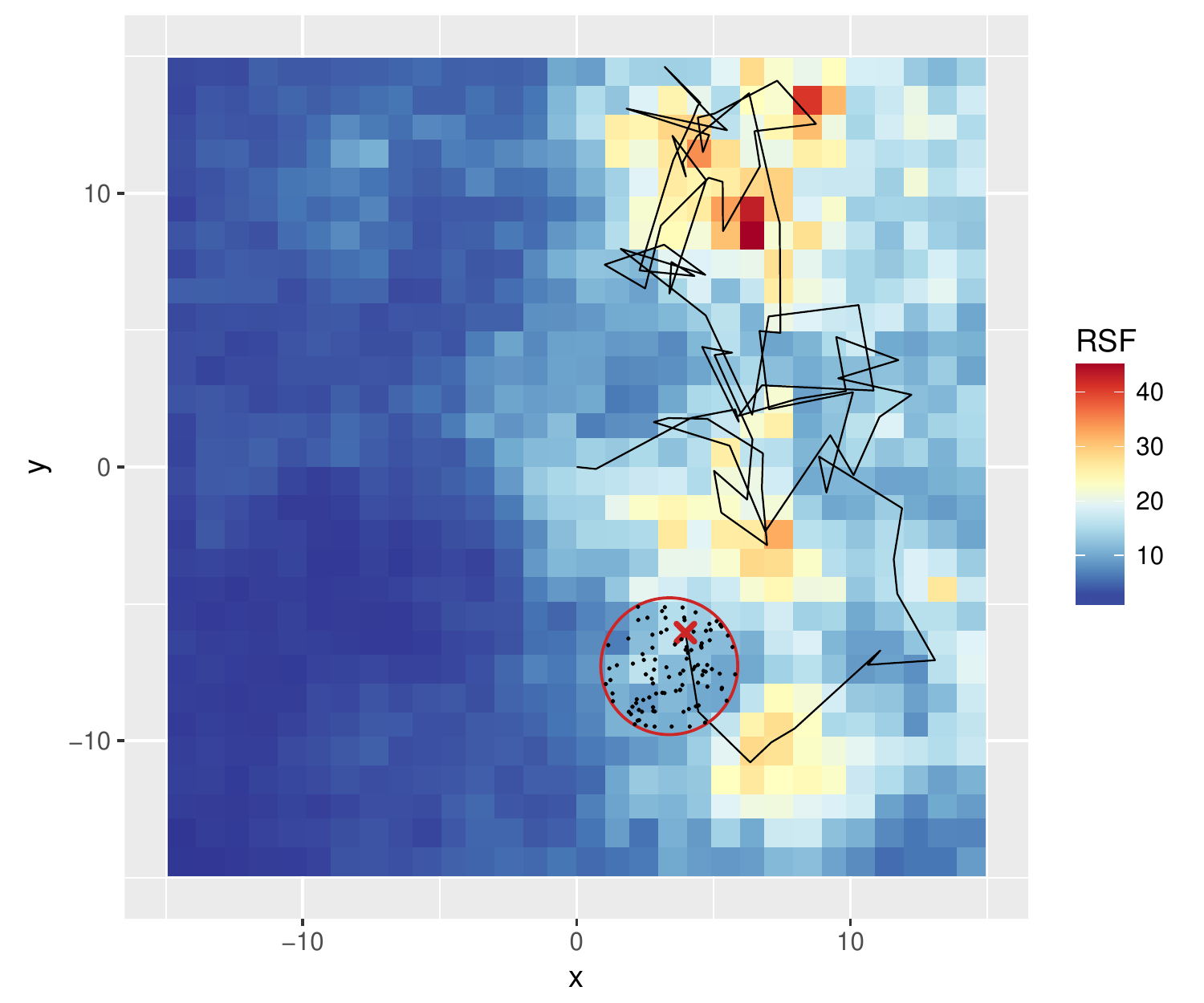}
  \caption{Illustration of the local Gibbs sampler in two dimensions. The background is the RSF; the solid line is the simulated track up to time $t$; the red cross is the current location $\bm{x}_t$; the red circle delimits $\mathcal{D}_r(\bm{c})$. The next location $\bm{x}_{t+1}$ is sampled from the black dots, with probabilities proportional to their RSF values.}
  \label{fig:gibbsillu}
\end{figure}

The local Gibbs algorithm, described in Section \ref{sec:gibbs}, can be used to simulate tracks based on a known RSF. The truncation of the RSF to the disc $\mathcal{D}_r(\bm{c})$ requires the calculation of the normalizing constant $C_r(\bm{c})$.
It is not generally possible to derive it analytically, but Monte Carlo sampling can be used to approximate it. In practice, to sample from the truncated target distribution $\tilde\pi$, $n_d$ points are generated uniformly in $\mathcal{D}_r(\bm{c})$, and $\bm{x}_{t+1}$ is sampled from those points, with probabilities proportional to their RSF values. Simulation using the local Gibbs algorithm is illustrated in Figure \ref{fig:gibbsillu}.

Here, we illustrate the method described in Section \ref{sec:method}, with the local Gibbs sampler. In Section \ref{sec:simLG}, we show that our algorithm can produce movement tracks on a known utilisation distribution and, in Section \ref{sec:estim}, we illustrate the use of the local Gibbs movement model for the estimation of resource selection and movement parameters from simulated data. The R code used for the simulations is available in the supplementary material (published with the manuscript in \emph{Ecology}).

\subsection{Simulated resources}
\label{sec:simrsf}
\begin{figure}[htbp]
  \centering
  \includegraphics[width=\textwidth]{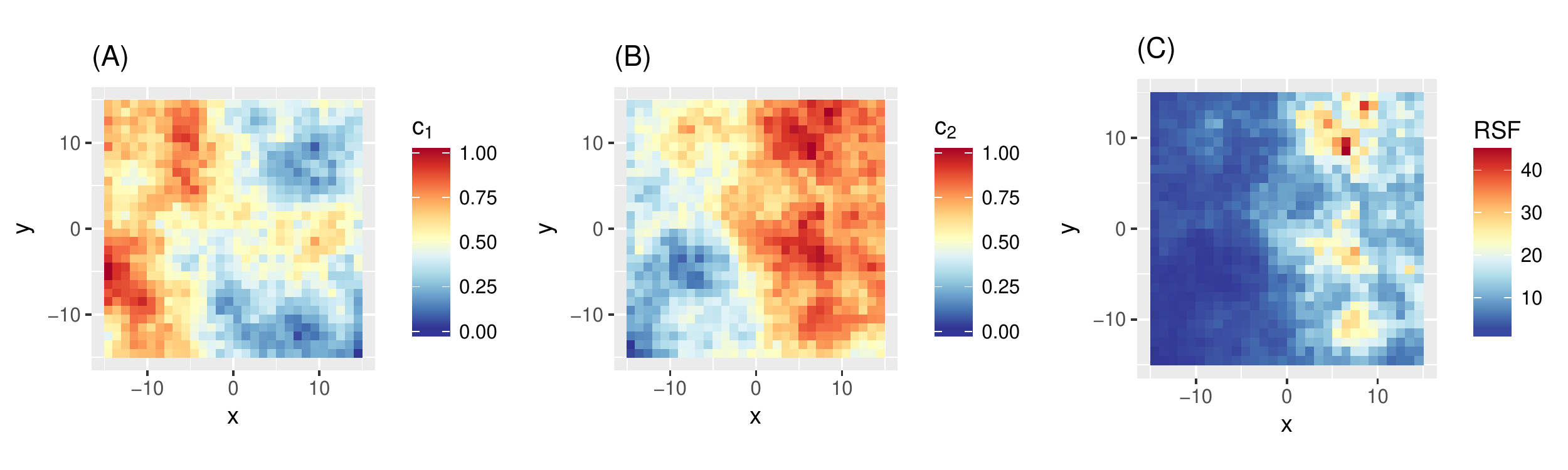}
  \caption{Resource distributions $r_1$ (A) and $r_2$ (B), and RSF (C), for the simulations.}
  \label{fig:simrsf}
\end{figure}

To mimic the type of environmental data of a real case study, we simulated two covariate distributions $c_1$ and $c_2$ as Gaussian random fields on square cells of size 1, using the R package gstat \citep{pebesma2004multivariable}. We restricted the study region to $\Omega = [-15,15] \times [-15,15]$, to ensure that the target distribution is integrable. Plots of $c_1$ and $c_2$ are shown in Figure \ref{fig:simrsf}(A) and \ref{fig:simrsf}(B). The utilisation distribution was defined by

\begin{equation*}
\pi(\bm{x}) = \dfrac{\exp( \beta_1 c_1 (\bm{x}) + \beta_2 c_2 (\bm{x}))}{\int_{\bm{z} \in \Omega} \exp( \beta_1 c_1 (\bm{z}) + \beta_2 c_2 (\bm{z})) d\bm{z}},
\end{equation*}
with $\beta_1 = -1$ and $\beta_2 = 4$ (i.e.\ avoidance for $c_1$ and preference for $c_2$). A plot of the RSF is shown in Figure \ref{fig:simrsf}(C).

\subsection{Local Gibbs simulation}
\label{sec:simLG}

In this section, we demonstrate that the local Gibbs algorithm, described in Section \ref{sec:gibbs}, can be used to sample from a given probability distribution. We considered the utilisation distribution $\pi$ defined in Section \ref{sec:simrsf}. To analyse the behaviour of the local Gibbs sampler at different spatial scales, we ran three simulations, with three different values for the radius $r$ of the movement kernel: $r=0.5$, $r=2$, and $r=8$. The value of $r$ affects the range of perception of the animal and, indirectly, its speed. For each $r$, $5 \times 10^5$ locations were simulated with the local Gibbs algorithm, starting from the point $\bm{x}_1=(0,0)$. (Given the length of the simulated tracks, the choice of the starting point has only a minor impact on the overall distribution of sampled locations.)

For comparison, we also illustrate the results of \cite{barnett2008analytic}, that the steady-state distribution of a standard SSF model ($\pi$ in Equation \ref{eqn:steadystate}) differs from the normalized SSF. We sampled a movement track from a step selection model with uniform sampling, as defined by \cite{forester2009accounting}, that we denote SSF$_{\text{unif}}$. We simulated $5 \times 10^5$ locations from SSF$_\text{unif}$, as follows. We started from $\bm{x}_1 = (0,0)$. Then, at each time step $t=1,2,\dots$, we generated 100 proposed locations $\{\bm{y}_1,\bm{y}_2,\dots,\bm{y}_{100}\}$ uniformly from a disc of radius $r=3$ centred on $\bm{x}_t$. The next location $\bm{x}_{t+1}$ was sampled from the proposed locations, with each point $\bm{y}_i$ having a probability to be picked proportional to $\pi(\bm{y}_i)$. That is, we use $\pi$ as the (normalized) SSF to simulate from the uniform sampling model. Here, we chose $r=3$ because it gave rise to approximately the same mean step length as the local Gibbs sampler with $r=2$ (i.e.\ comparable speed of spatial exploration).

\begin{figure}[htbp]
  \centering
  \includegraphics[width=0.85\textwidth]{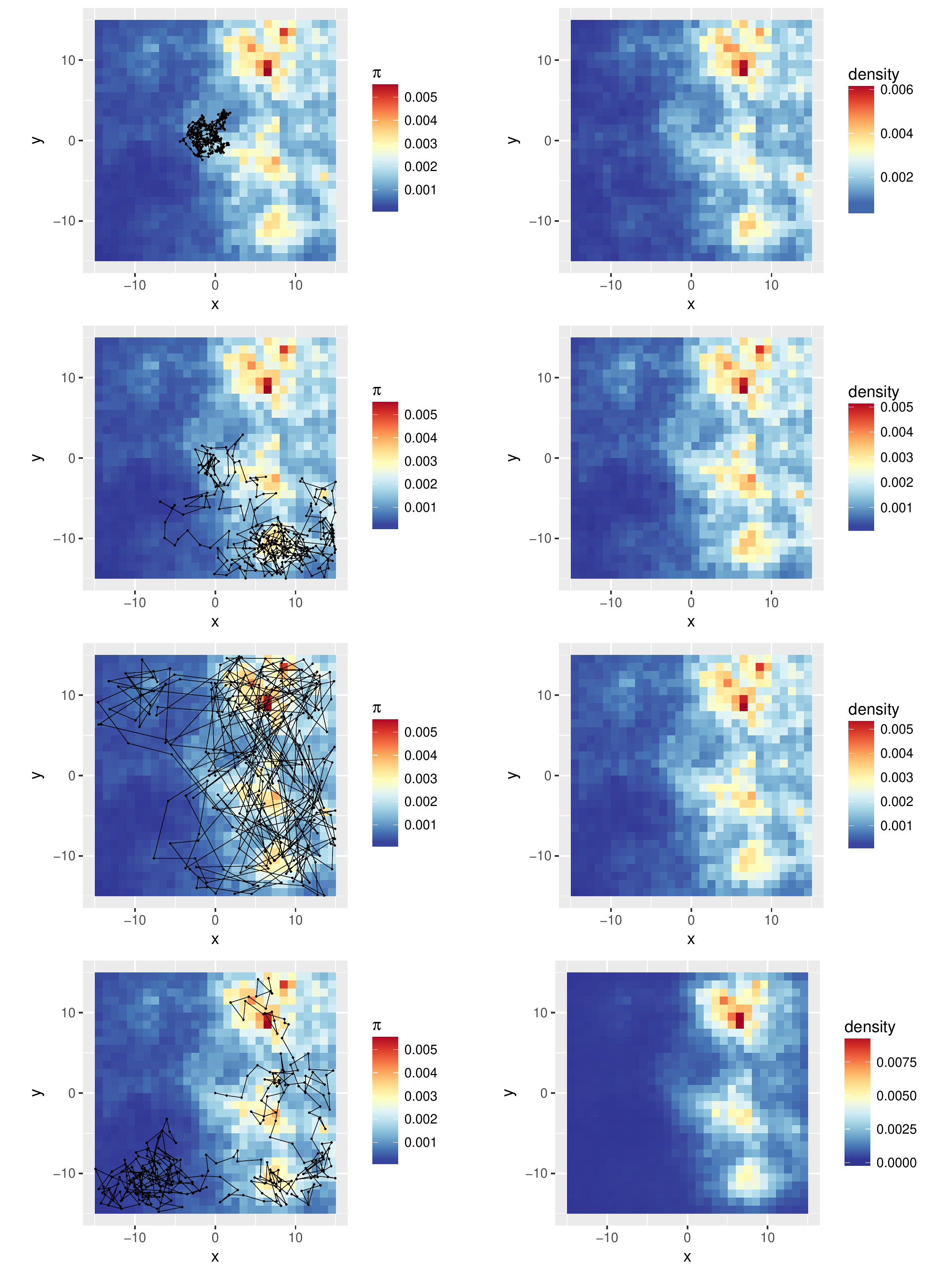}
  \caption{Simulation using a local Gibbs sampler, with radius parameter $r=0.5$ (first row), $r=2$ (second row), and $r=8$ (third row); and simulation using a step selection function with uniform sampling (r=$3$, fourth row). The left column displays the first 300 simulated steps, and the background colour represents the utilisation distribution (i.e.\ the normalized RSF; the RSF is given in Figure \ref{fig:simrsf}(C)). The right column shows the density of the $5 \times 10^5$ simulated locations, i.e.\ the normalized counts.}
    \label{fig:simtrack1}
\end{figure}

The first 300 steps of each simulated track, and the density of all simulated points, are shown in Figure \ref{fig:simtrack1}. The density of points simulated from the local Gibbs sampler (right column, first three plots) displays the same patterns as the true RSF (Figure \ref{fig:simrsf}(C)). By contrast, the density of the locations obtained in the SSF$_{\text{unif}}$ simulation (right column, last plot) fails to capture many features of the landscape, as the process spends a disproportionate amount of time in areas of high values of $w(\bm{x})$.

\begin{figure}[htbp]
  \centering	
  \includegraphics[width=0.7\textwidth]{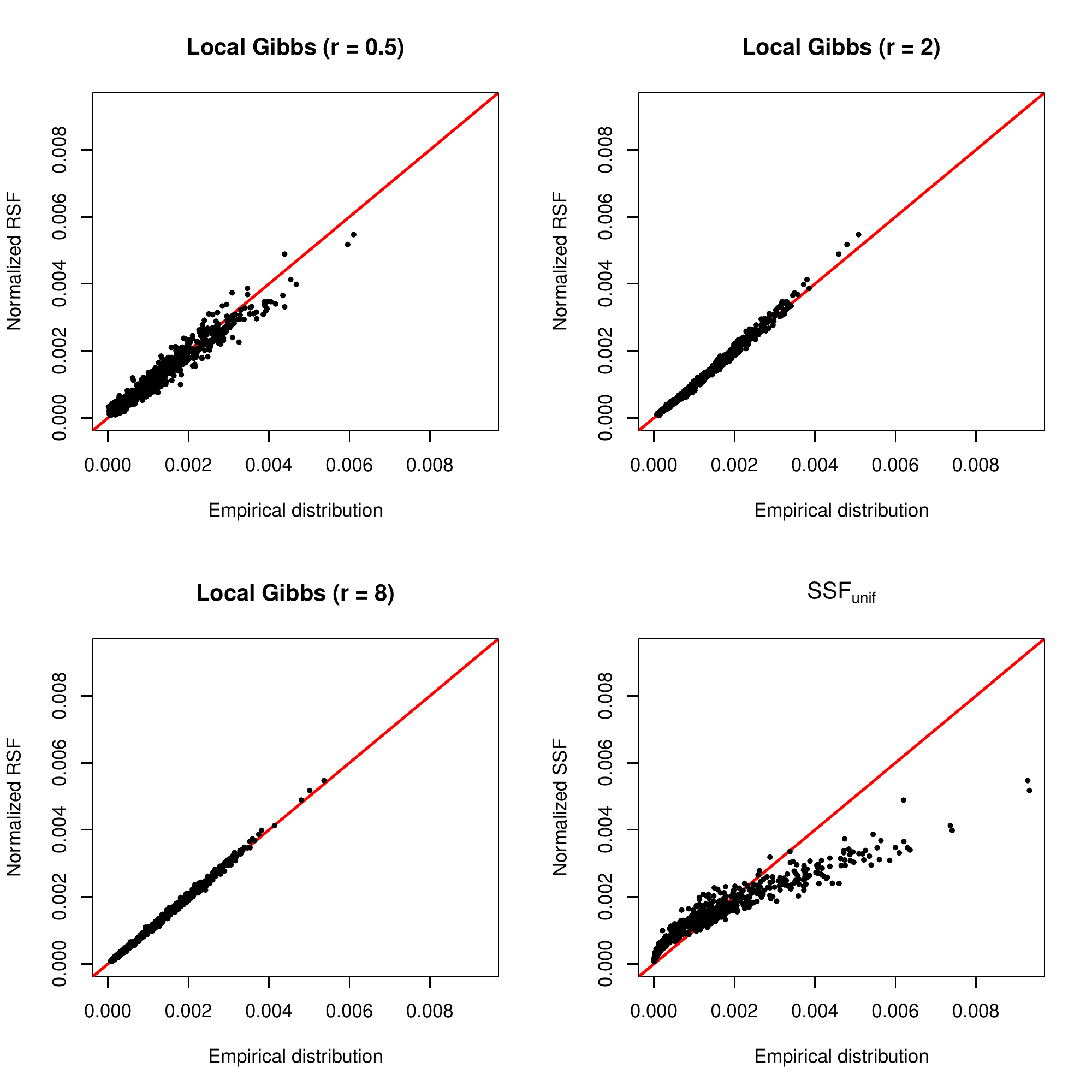}
  \caption{Results of the simulations. In each plot, the distribution of simulated points (on the x-axis) is compared to the distribution $\pi$ used in the simulations (on the y-axis). In the local Gibbs simulations, $\pi$ denotes the (normalized) RSF and, in the SSF$_\text{unif}$ simulation, $\pi$ denotes the (normalized) SSF. Each dot represents the value associated with one grid cell. The closer the dots are to the identity line, the more similar the empirical distribution is to $\pi$. In the local Gibbs simulations, the empirical distributions are very similar to the RSF; the similarity increases with $r$, because a larger radius leads to faster spatial exploration. For the SSF$_\text{unif}$ model, there is a clear discrepancy between the empirical distribution and the SSF, as predicted by \cite{barnett2008analytic} (Equation 4).}
    \label{fig:counts_vs_true}
\end{figure}

To compare the empirical distribution of simulated points to the distribution $\pi$ used in the simulations, we plotted the (normalized) count of locations simulated in each grid cell against the corresponding value of $\pi$. The comparison is presented in Figure \ref{fig:counts_vs_true}. Alignment with the identity line indicates similarity between the empirical distribution and $\pi$. For the three local Gibbs simulations, the points align well with the identity line -- in particular in the experiments with $r=2$ and $r=8$, in which the speed of spatial exploration is higher than when $r=0.5$. This confirms that the local Gibbs algorithm can sample movement trajectories on a given target distribution. It defines a movement model for which the long-term distribution of locations is known. However, the plot for the SSF$_\text{unif}$ simulation reveals a clearly non-linear relationship between the density of simulated points and the normalized SSF. This confirms the results of \cite{barnett2008analytic}, \cite{avgar2016integrated}, and \cite{signer2017estimating}: the coefficients of a step selection function do not measure the underlying steady-state distribution. (Note that SSF models may be used to estimate space use, with simulations, as in \cite{avgar2016integrated}, but the parameters of the SSF only measure local habitat selection.) We illustrated how the local Gibbs sampler can generate movement tracks that converge in distribution to the underlying RSF.

\subsection{Local Gibbs estimation}
\label{sec:estim}
The approach introduced in Section \ref{sec:method} shows great promise for the estimation of movement and resource selection parameters from observed animal movement data. Considering the MCMC algorithm as a movement model, it is in principle straightforward to express the likelihood of observed steps, given the parameters of the sampler (e.g.\ radius $r$ in the local Gibbs model) and of the RSF ($\beta_1,\beta_2,\dots$). In cases where the transition kernel of the chosen sampler, $p(\bm{x}_{t+1} \vert \bm{x}_t)$, can be calculated, the likelihood of $T$ observations $(\bm{x}_1, \bm{x}_2, \dots, \bm{x}_T)$ is derived as $L = \prod_{t=1}^{T-1} p(\bm{x}_{t+1} \vert \bm{x}_t)$. 

In this section, we wish to demonstrate its practical application, with the example of the local Gibbs model. We simulated a track of $T=3000$ locations from the local Gibbs sampler (described by the algorithm in Section \ref{sec:gibbs}), with $r=2$, on the RSF defined in Section \ref{sec:simrsf}. Then, similarly to a real analysis, we used the local Gibbs model to recover estimates of the RSF (i.e.\ of $\beta_1$ and $\beta_2$) and of $r$, from the (simulated) movement data and covariate rasters.

The likelihood of an observed track under the local Gibbs model is obtained as the product of the likelihoods of the individual steps,

\begin{equation} \label{eqn:LGlk}
  L = \prod_{t=1}^{T-1} p(\bm{x}_{t+1} \vert \bm{x}_t) =
  \prod_{t=2}^T \dfrac{1}{\pi r^2} 
  \int_{\bm{c} \in \mathcal{D}_r(\bm{x}_t) \cap \mathcal{D}_r(\bm{x}_{t+1})}
  \dfrac{\pi(\bm{x}_{t+1})}{\int_{\bm{z} \in \mathcal{D}_r(\bm{c})} \pi(\bm{z}) d\bm{z}} d\bm{c}
\end{equation}

The details of the derivation are given in Appendix S1. This likelihood is a function of the movement parameter $r$, and of the coefficients $\beta_i$ of the RSF (which appear in the expression of $\pi$). Maximum likelihood techniques can then be used to obtain parameter estimates. We implemented the likelihood function of Equation \ref{eqn:LGlk}, and used the numerical optimiser \texttt{nlminb}, in R, to get maximum likelihood estimates of $\beta_1$, $\beta_2$, and $r$. The results are summarized in Table \ref{tab:MLEs}.

\begin{table}[htbp]
  \centering
  \begin{tabular}{lccc}
    \toprule
    Parameter & True value & Estimate & 95\% confidence interval\\
    \midrule
    $\beta_1$ & -1 & -0.86 & [-1.46,-0.26] \\
    $\beta_2$ & 4 & 4.15 & [3.53,4.77] \\
    $r$ & 2 & 2 & [1.81,2.20] \\
    \bottomrule
  \end{tabular}
  \caption{Maximum likelihood estimates and Hessian-based 95\% confidence intervals for the parameters of the local Gibbs model, obtained for one simulated track.}
  \label{tab:MLEs}
\end{table}

\begin{figure}[htbp]
  \centering
  \includegraphics[width=0.5\textwidth]{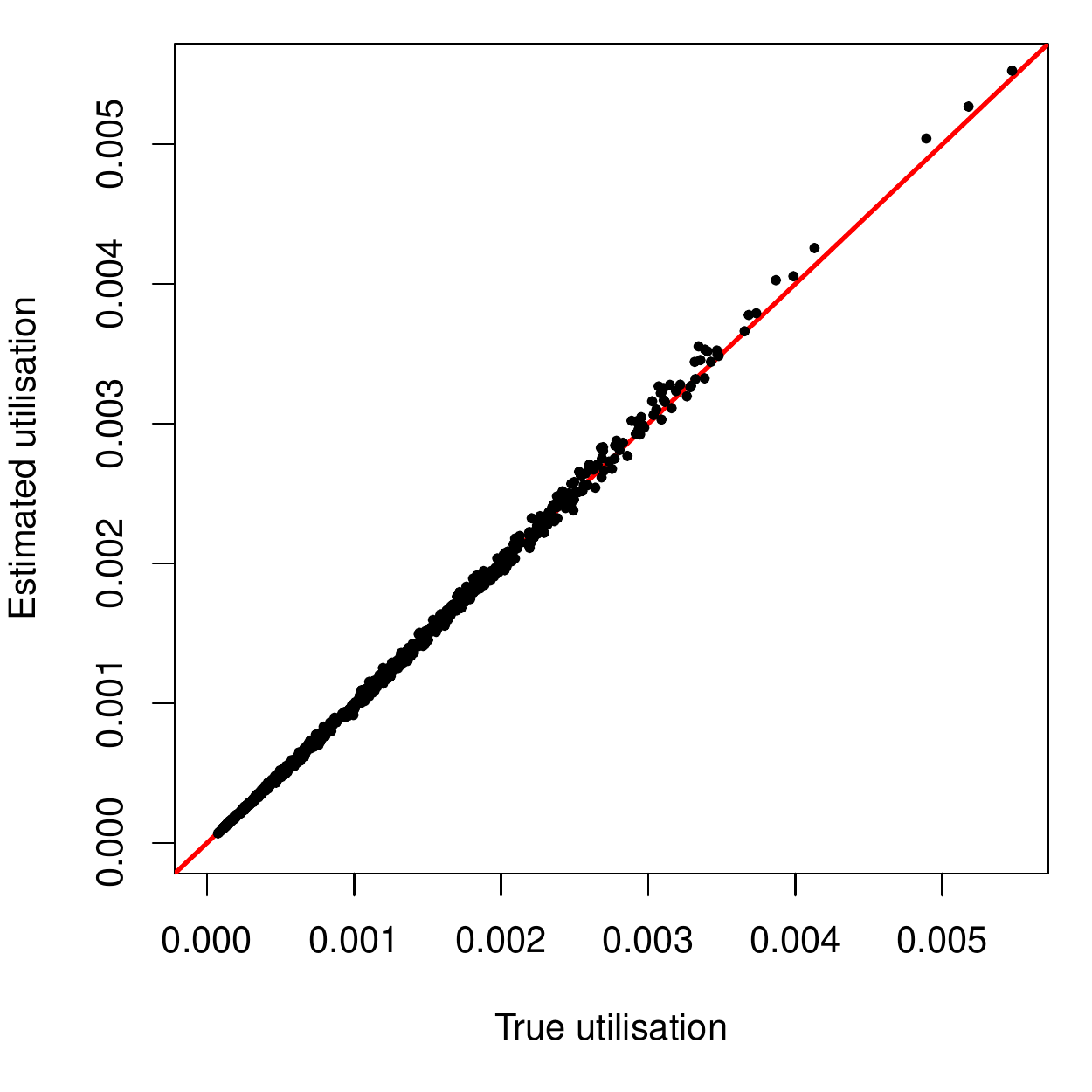}
  \caption{Utilisation estimates obtained for a simulated track, using the local Gibbs model. Each dot corresponds to one grid cell. The x axis shows the value of the true utilisation distribution, and the y axis shows the value of the estimated utilisation distribution, for each grid cell.}
  \label{fig:estimUD}
\end{figure}

Figure \ref{fig:estimUD} shows a plot of the estimated utilisation value of each grid cell against its true utilisation value. If we denote $\hat{w}_i$ the estimated value of the RSF in cell $i$, its estimated utilisation value $\hat{\pi}_i$ is derived as
\begin{equation*}
  \hat{\pi}_i = \dfrac{\hat{w}_i}{\sum_{j \in \text{cells}} \hat{w}_j}.
\end{equation*}

In Figure \ref{fig:estimUD}, the alignment of the dots with the identity line indicates that the estimated utilisation distribution captures the shape of the true utilisation distribution well. In addition, the parameter $r$ of the movement process was successfully estimated (Table \ref{tab:MLEs}).

This example demonstrates how the method can be used to estimate resource selection and movement parameters from tracking data. In real applications, unlike with simulated data, the true form of the movement process would not be known, and additional work would be needed to assess the fit. We discuss this further in Section \ref{sec:discussion}.

\section{Discussion}
\label{sec:discussion}
We have presented a versatile class of models of animal movement, for which the steady-state distribution of locations is proportional to the same resource selection function that influences short-term movement. Our approach reconciles the resource selection and step selection approaches to the analysis of space use data. We anticipate that the resolution of this discrepancy between RSF and SSF models will have important implications for the study of individual movement and, also, species distributions. The central point of this paper is the idea that multiscale modelling of a dynamic system can be achieved using stochastic processes for which both the short-term transition density and the long-term stationary distribution are explicitly formulated (in particular, here, MCMC samplers). Although we have presented this method for the analysis of animal movement and resource selection, we expect that the underlying idea could have other ecological applications. For example, this problem is reminiscent of population genetics, where both the microscopic heritability laws and the macroscopic allele frequencies are of interest.

At the level of the individual, we have recognised a tendency in the current literature to embed increasingly realistic movement models in SSF analyses. We hazard that the subtext of this trend is the intuitive notion that the habitat selection coefficients of SSF models that stay faithful to movement biology, will automatically correspond to the estimates of RSF models. As we have argued and demonstrated here, this is not necessarily the case, because SSF coefficients measure local habitat selection rather than long-term space use. Conversely, any given population distribution may be achievable by multiple movement models -- just as, in the simplest of movement models, the same degree of population diffusivity can be achieved by an infinity of different movement rules, simply by trading off individual speed against path sinuosity. Although meticulous realism in movement turns out not to be a strict requirement for achieving agreement between the microscopic and macroscopic models of space use, our paper demonstrates how SSFs (through the application of statistical estimation and model selection) might in the future be used to learn about movement biology.

This manuscript serves as a proof of concept for the approach, but stops short of describing a complete workflow for the analysis of animal location data. In Section \ref{sec:estim}, using simulated data, we explained how the local Gibbs model can be used to estimate resource selection and movement parameters from a movement track. In a real data analysis, it would be necessary to investigate the goodness-of-fit. One possibility would be to simulate many locations from the fitted local Gibbs sampler, and compare the simulated and observed data in terms of some metrics of movement (e.g.\ distribution of step lengths). Discrepancies between features of the true and simulated data sets would point to possible model misspecifications. In addition, different models of individual movement, described by different MCMC algorithms but all guaranteed to scale up to the same long-term distribution, may be allowed to compete in a setting of statistical model selection, pointing to parsimonious explanations of the movement observations. In the estimation framework introduced in Section \ref{sec:estim}, likelihood-based model selection criteria, such as the AIC, could be used to compare several candidate models. The likelihood derived from a MCMC movement model accounts for the serial correlation found in telemetry data. As such, it is a more defensible measure of likelihood than what might be obtained with other RSF approaches \citep{aarts2008estimating, fieberg2010correlation}.

This modelling framework combines some of the advantages of process-based movement models and of distribution-based resource selection models. In addition to its advantages for individual-level inference, the prospect of reconciliation between RSF and SSF approaches will also benefit population-level results. In particular, the problem of formally combining the two major sources of space-use information -- telemetry and transect data -- has, in our experience, resisted several analytical attempts. The approach proposed here offers a solution to this problem of joint inference. For example, the steady-state distribution implied by a SSF fitted to telemetry data would be required to coincide with the utilisation distribution generated by fitting a RSF to independently obtained transect data. As described in Section \ref{sec:estim}, the likelihood of a track $(\bm{x}_1, \dots ,\bm{x}_T)$ under a MCMC movement model with transition kernel $p(\bm{x}_{t+1} \vert \bm{x}_t)$ is $L_\text{mov} = \prod_{t=1}^{T-1} p(\bm{x}_{t+1} \vert \bm{x}_t)$ and, in the same framework, the likelihood $L_\text{ind}$ of isolated survey locations $\{ \bm{y}_1, \dots, \bm{y}_n \}$ can be obtained using standard RSF methods (e.g.\ logistic regression or Poisson GLM). The two types of data can be combined by multiplying $L_\text{mov}$ and $L_\text{ind}$, thus enhancing the effective sample size of the resulting estimates. Incorporating additional constraints, for example if the survey is confined to a subregion, is also straightforward.

Because it builds on the very wide and flexible class of MCMC samplers, various other movement rules could be considered. The slice sampler \citep{neal2003} is an existing rejection-free sampler that shares some mathematical details with our local Gibbs sampler, and a `local' version may give some additional flexibility in movement modelling.  Models of animal movement often incorporate directional persistence, such as the discrete-time and continuous-time correlated random walks \citep[e.g.,][respectively]{jonsen2005robust, johnson2008continuous}. Within the framework we described, this feature of movement could be modelled using non-reversible MCMC samplers, which often display this type of autocorrelation \citep[e.g.][]{michel2017forward}. Such algorithms could be used for more realistic movement models. 

Although we have focused on the case where the radius parameter $r$ of the local Gibbs algorithm is taken to be constant, allowing $r$ to be stochastic is straightforward, as mentioned above. The flexibility of the model depends in part on the choice of this distribution. More realistic features of animal movement, such as different distributions of step lengths, could thus be incorporated in the local Gibbs sampler by choosing a flexible parametric distribution for $r$ (e.g.\ a gamma or Weibull distribution).
A further refinement would be to combine this approach with the state-space modelling framework \citep{patterson2008state}, with the state of the process representing true location, thus incorporating measurement error on locations and giving some robustness against errors of measurement, classification or registration in the habitat map.
    
The present paper therefore opens the way for future research in three vital directions: the exploration of the wealth of biological models that can be implemented with our MCMC analogues, the development of inferential methods for the integrated analysis of different data types, and the investigation into how population-level space use arises from individual rules of movement.

\subsection*{Acknowledgements}
TM was supported by the Centre for Advanced Biological Modelling at the University of Sheffield, funded by the Leverhulme Trust, award number DS-2014-081. We thank Jonathan Potts for fruitful discussions on an earlier version of the manuscript, and the associate editor and two reviewers whose suggestions greatly improved the presentation of this work.

\bibliographystyle{apalike}
\bibliography{refs.bib}

\newpage
\section*{Appendix S1. Movement kernel for the local Gibbs model}
For a fixed radius $r$ and any target distribution $\pi$, we have

\begin{align*}
  p(\bm{x}_{t+1} \vert \bm{x}_{t}) & = \int_{\bm{c} \in \mathcal{D}_r(\bm{x}_t)} p(\bm{x}_{t+1} \vert \bm{c})
                                     p(\bm{c} \vert \bm{x}_t) d\bm{c}\\
                                   & = \int_{\bm{c} \in \mathcal{D}_r(\bm{x}_t)} 
                                     \dfrac{\pi(\bm{x}_{t+1}) I_{\{\bm{x}_{t+1}\in \mathcal{D}_r(\bm{c})\}}}
                                     {\int_{\bm{z} \in \mathcal{D}_r(\bm{c})} \pi(\bm{z}) d\bm{z}} \dfrac{1}{\pi r^2} d\bm{c}\\
                                   & = \dfrac{1}{\pi r^2} \int_{\bm{c} \in \mathcal{D}_r^{(t)}}
                                     \dfrac{\pi(\bm{x}_{t+1})}{\int_{\bm{z} \in \mathcal{D}_r(\bm{c})} \pi(\bm{z}) d\bm{z}} d\bm{c},
\end{align*}
where $\mathcal{D}_r^{(t)} = \mathcal{D}_r(x_t) \cap \mathcal{D}_r(x_{t+1})$, and $I$ is the indicator function. Then, if the target distribution is flat, say 
\begin{equation*}
  \forall \bm{x},\ \pi(\bm{x})=k,
\end{equation*}
we have
\begin{align*}
  p(\bm{x}_{t+1} \vert \bm{x}_{t}) & = \dfrac{1}{\pi r^2} \int_{\bm{c} \in \mathcal{D}_r^{(t)}} 
                                     \dfrac{k}{\int_{\bm{z} \in \mathcal{D}_r(\bm{c})} k\ d\bm{z}} d\bm{c}\\
                                   & = \dfrac{k}{\pi r^2} \int_{\bm{c} \in \mathcal{D}_r^{(t)}} 
                                     \dfrac{1}{k \pi r^2} d\bm{c} \\
                                   & = \dfrac{1}{(\pi r^2)^2} \int_{\bm{c} \in \mathcal{D}_r^{(t)}} d\bm{c}\\
                                   & = \dfrac{1}{(\pi r^2)^2} \mathcal{A}(\mathcal{D}_r^{(t)})
\end{align*}
where $\mathcal{A}(\mathcal{D}_r^{(t)})$ is the area of $\mathcal{D}_r^{(t)}$. It can be shown that
\begin{equation*}
  \mathcal{A}(\mathcal{D}_r^{(t)}) = 
  \begin{cases}
    2 r^2 \cos^{-1} \left( \dfrac{d_t}{2r} \right) - d_t \sqrt{r^2 - \dfrac{d_t^2}{4}} & 
    \text{ if } d_t < 2r,\\
    0 & \text{otherwise,}
  \end{cases}
\end{equation*}
where $d_t = \lVert \bm{x}_{t+1} - \bm{x}_t \rVert$ is the distance between $\bm{x}_t$ and $\bm{x}_{t+1}$.

\end{document}